\documentstyle[12pt,epsfig]{article}

\newcommand{\be}{\begin{equation}}
\newcommand{\ee}{\end{equation}}
\newcommand{\Dlt}{\Delta}
\newcommand{\dlt}{\delta}

\newcommand{\br}{{\bf r}}
\newcommand{\bk}{{\bf k}}

\newcommand{\ba}{{\bf a}}

\newcommand{\bt}{\beta}
\newcommand{\vp}{\varphi}
\newcommand{\ep}{\varepsilon}
\newcommand{\al}{\alpha}
\newcommand{\ra}{\rightarrow}
\newcommand{\sgm}{\sigma}

\newcommand{\gm}{\gamma}
\newcommand{\om}{\omega}
\newcommand{\Om}{\Omega}

\newcommand{\dgr}{\dagger}
\newcommand{\lbd}{\lambda}

\begin{document}

\begin{center}

{\Large {\bf Cold atoms in double-well optical
lattices} \\ [5mm]

V.I. Yukalov$^1$ and E.P. Yukalova$^2$} \\ [3mm]

{\it
$^1$Bogolubov Laboratory of Theoretical Physics, \\
Joint Institute for Nuclear Research, Dubna 141980, Russia \\ [3mm]

$^2$Department of Computational Physics, Laboratory of Information
Technologies,\\
Joint Institute for Nuclear Research, Dubna 141980, Russia}

\end{center}

\begin{abstract}

Cold atoms, loaded into an optical lattice with double-well sites,
are considered. Pseudospin representation for an effective Hamiltonian
is derived. The system in equilibrium displays two phases, ordered and
disordered. The second-order phase transition between the phases can
be driven either by temperature or by changing the system parameters.
Collective pseudospin excitations have a gap disappearing at the
phase-transition point. Dynamics of atoms is studied, when they are
loaded into the lattice in an initially nonequilibrium state. It is
shown that the temporal evolution of atoms, contrary to their equilibrium
thermodynamics, cannot be described in the mean-field approximation, since
it results in  a structurally unstable dynamical system, but a more accurate
description is necessary taking account of attenuation effects.

\end{abstract}

\vskip 1cm

{\bf PACS}: 03.75.Lm, 03.75.Be, 03.67.Lx

\vskip 1cm

\newpage

\section{Introduction}

Degenerate cold gases, Bose [1--8] as well as Fermi [9,10], possess
many interesting properties. Loading cold atoms into optical lattices
yields highly controllable systems that could be employed for a variety
of applications [11--15]. Recently, a novel type of optical lattices
has been experimentally realized, each site of which is formed by a
double-well potential [16--21].

The phase diagram of cold bosons in double-well optical lattices has 
been studied in Refs. [22,23], where the Hubbard model is used and the 
main attention is payed to the peculiarities of the superfluid-insulator 
phase transition in such lattices, as compared to this transition in the 
standard single-well optical lattices [11--14]. In Ref. [24], the 
superfluid phase is studied, aiming at finding a type of superfluid 
with the broken time-reversal symmetry.

The aim of the present paper is to study a different regime that can 
be realized in the double-well optical lattices. We consider the 
lattices that are in the insulating state, far from the boundary of 
the insulator-superfluid transition. In that state, the jumps of atoms 
between different lattice sites are suppressed, while the tunneling 
between the wells of a single-site double well can be of principal 
importance. We show that in such insulating double-well optical 
lattices there exists another phase transition, the {\it order-disorder
phase transition}. The investigation of this different physical regime 
in insulating double-well optical lattices, far from the superfluid 
phase-transition, but close to another, order-disorder phase transition, 
is the main motivation for the present work.

The order-disorder phase transition, and the related ordered and 
disordered states, can be the most clearly realized for the case of 
one atom per a double well. Therefore, we consider exactly this case. 
An additional argument for considering the lattices with one atom per 
site is that the optical lattices with small filing factors, such as 
one or two, seem to be good candidates for quantum information 
processing [11,19].

The advantage of dealing with the case of one atom per a double well 
is twofold. The most important, as is emphasized above, is that this is 
the setup allowing for a clear realization of the order-disorder phase 
transition. Another, technical, convenience is that in this case the 
pseudospin representation can be invoked. Some types of spin Hamiltonians 
using cold atoms in optical lattices can be met in literature. It is 
straightforward to obtain a spin Hamiltonian for {\it spinor condensates} 
[25]. The Hubbard model for a system of {\it two-component bosons} can be 
reduced to a pseudospin representation by a second-order perturbation 
theory in the tunneling parameter [26--28]. In our case of an insulating 
double-well optical lattice, the pseudospin representation can be 
introduced without perturbation theory, by means of an exact canonical 
transformation, similar to the canonical transformations used for deriving 
the spin representations for superconductors and ferromagnets [29,30].

The key point of realizing the ordered and disordered states in a 
double-well optical lattice is the existence of sufficiently long-range 
atomic interactions. Such interactions of dypolar type arise between polar
 molecules [31], Rydberg atoms [32], and between atoms with large magnetic 
moments [33]. The existence of long-range atomic interactions is another 
point making our consideration principally different from the earlier 
theoretical works [22--24] on the double-well lattices.

Characterizing collective atomic states in the double-well lattices, we, 
first, describe their equilibrium properties. However, we keep in mind that 
in experiments atoms need to be loaded into a lattice, and their initial 
state after the loading may be nonequilibrium. If so, how then atoms would 
relax to their equilibrium state? Another way, when atoms can happen to be 
in a nonequilibrium states, is if the lattice parameters are varied after 
the atoms have been loaded. If this variation is sufficiently fast, atoms 
again occur to be in a nonequilibrium state, from which they should relax to 
an equilibrium one. In order that our consideration of the insulating 
double-well optical lattices would be more complete, and keeping in mind 
the possibility of realizing nonequilibrium states, we study not only 
the equilibrium properties of such lattices, but also the relaxational 
dynamics of atoms from initially nonequilibrium states to their equilibrium.

The paper is organized as follows. First, we derive an effective
Hamiltonian for the considered insulating system and show that it allows
for a convenient pseudospin representation (Sec. II). Then we study the
equilibrium thermodynamics of the model, which exhibits the existence
of two phases, ordered and disordered (Sec. III). Collective excitations,
corresponding to pseudospin waves, are described in Sec. IV. Dynamics
of atoms, loaded in the lattice in an initially nonequilibrium state,
is considered in Sec. V. Lyapunov stability and structural stability of
solutions to the equations of motion are investigated in Sec. VI, where
the principal importance of taking into account attenuation effects is
demonstrated. The main results are summarized in Sec. VII.

\section{Model Hamiltonian}

We start with the general form of the energy Hamiltonian
\be
\label{1}
\hat H = \int \psi^\dgr(\br) H_L(\br) \psi(\br) \; d\br \; +
\; \frac{1}{2}\; \int \psi^\dgr(\br) \psi^\dgr(\br')
\Phi(\br-\br') \psi(\br') \psi(\br) \; d\br d\br' \; ,
\ee
where $\psi(\br)=\psi(\br,t)$ is a field operator, in which the time
dependence, for brevity, is omitted. Keeping in mind the case of an
insulating lattice, with the unity filling factor, the statistics of
atoms is not as important. So, atoms can be either bosons or fermions.
In the lattice Hamiltonian
\be
\label{2}
H_L(\br) \equiv -\; \frac{\nabla^2}{2m} \; + \; V(\br) \; ,
\ee
the lattice potential $V(\br+\ba_i)=V(\br)$ is periodic over the lattice
$\{\ba_i : \; i=1,2,\ldots,N_L\}$ and enjoys the double-well structure
around each of the lattice sites $\ba_i$. The interaction potential
$\Phi(-\br)=\Phi(\br)$, in general, is a sum of the short-range
interactions, whose strength can be regulated by the Feschbach resonance
techniques [2,10,15,34,35], and of a long-range interaction of the dipolar
type, such that exists between polar molecules [31], Rydberg atoms [32],
and atoms with large magnetic moments [33]. If, instead of atoms, we consider
ions, then these exists the long-range Coulomb interaction. The strength of
the interaction potential is assumed to be such that the intersite interactions,
at least between the nearest neighbors, cannot be neglected.

The field operator can be expanded over Wannier functions
\be
\label{3}
\psi(\br) =\sum_{nj} c_{nj} w_n(\br-\ba_j) \; ,
\ee
where $n$ is a band index and $j$ enumerates the lattice sites. Then
Hamiltonian (1) transforms into
\be
\label{4}
\hat H = \sum_{ij} \; \sum_{mn}
E_{ij}^{mn} c_{mi}^\dgr c_{nj} \; + \;
\frac{1}{2} \; \sum_{ \{ j\} } \;
\sum_{ \{ n\} } \; \Phi_{j_1j_2j_3j_4}^{n_1n_2n_3n_4} \;
c_{n_1j_1}^\dgr c_{n_2j_2}^\dgr c_{n_3j_3} c_{n_4j_4} \; ,
\ee
where
\be
\label{5}
E_{ij}^{mn} \equiv \int
w_m^*(\br-\ba_i) \; H_L(\br) \; w_n(\br-\ba_j) \; d\br
\ee
and $\Phi_{j_1j_2j_3j_4}^{n_1n_2n_3n_4}$ is the corresponding matrix
element of the interaction potential.

It can be shown (see Appendix A), that
\be
\label{6}
E_{ij}^{mn} = \dlt_{mn} E_{ij}^n
\ee
is diagonal with respect to the band indices. Here,
\be
\label{7}
E_{ij}^n = \dlt_{ij} E_n + ( 1 - \dlt_{ij} ) J_{ij}^n \; ,
\ee
with
$$
E_n = \int w_n^*(\br) \; H_L(\br)\; w_n(\br) \; d\br \; ,
\qquad J_{ij}^n = \int
w_n^*(\br-\ba_{ij}) \; H_L(\br)\; w_n(\br) \; d\br \; .
$$
Keeping in mind an insulating lattice implies that the intersite hopping
is small, such that
\be
\label{8}
\left | \frac{J_{ij}}{E_n} \right | \; \ll \; 1 \qquad (i\neq j) \; .
\ee

Assuming that each lattice site contains just one atom, we impose the
unipolarity conditions
\be
\label{9}
\sum_n c_{nj}^\dgr c_{nj} = 1 \; , \qquad c_{nj} c_{nj} = 0 \; .
\ee

Taking into account the above consideration, for Hamiltonian (4) we have
\be
\label{10}
\hat H = \sum_{nj} E_n c_{nj}^\dgr c_{nj} \; + \;
\frac{1}{2} \; \sum_{i\neq j} \;
\sum_{mnm'n'} \; V_{ij}^{mnm'n'} \;
c_{mi}^\dgr c_{nj}^\dgr c_{m'j} c_{n'i} \; ,
\ee
where
\be
\label{11}
V_{ij}^{mnm'n'} \equiv \Phi_{ijji}^{mnm'n'} \; \pm \;
\Phi_{ijij}^{mnn'm'} \; ,
\ee
with the upper or lower sign for bosons or fermions, respectively.

When a single-well potential is transformed into a double-well potential,
then each spectrum level of a particle, it would possess in the former
potential, splits into two lines for the particle in a double-well potential.
The spectrum splitting is connected with the particle tunneling between the
wells of a double-well potential. The splitting magnitude depends on the
characteristics of the double-well potential and can be regulated in a wide
range [36]. Thus, for a double-well lattice, one cannot limit oneself by
considering solely the lowest energy band, but at least two energy levels
must be taken into account, describing the level splitting and interwell
tunneling. In what follows, we take into account two lowest energy levels,
so that the index $n$ takes two values $n=1,2$.

The symmetry properties of the ground-state wave function and of that for
the first excited state are known [36] to be different. Enumerating the
ground state with $n=1$ and the excited state with $n=2$, one has
\be
\label{12}
w_1(-\br) = w_1(\br) \; , \qquad w_2(-\br) = - w_2(\br) \; .
\ee
Also, both functions $w_n(\br)$, for $n=1,2$, can be taken to be real.
Because of the symmetry property (12), the matrix elements of the type as
$V_{ij}^{1112}$ and $V_{ij}^{2221}$ become zero.

For what follows, it is convenient to introduce the notation
\be
\label{13}
E_0 \equiv \frac{1}{2} \; ( E_1 + E_2 ) \; .
\ee
And let us define the interaction matrix elements
$$
A_{ij} \equiv \frac{1}{4} \left ( V_{ij}^{1111} + V_{ij}^{2222} +
2 V_{ij}^{1221} \right ) \; , \qquad
B_{ij} \equiv \frac{1}{2} \left ( V_{ij}^{1111} + V_{ij}^{2222} -
2 V_{ij}^{1221} \right ) \; ,
$$
\be
\label{14}
C_{ij} \equiv \frac{1}{2} \left (  V_{ij}^{2222} - V_{ij}^{1111}
\right ) \; , \qquad I_{ij} \equiv - 2V_{ij}^{1122} \; .
\ee
The quantity
\be
\label{15}
\Om \equiv E_2 - E_1 + \sum_{j(\neq i)} C_{ij}
\ee
is the tunneling frequency characterizing the tunneling between the wells
of a double-well potential.

The convenience of dealing with the two-level case is that it allows for
the introduction of the pseudospin representation. The pseudospin operators
can be defined as
$$
S_j^x = \frac{1}{2} \left ( c_{1j}^\dgr c_{1j} -
c_{2j}^\dgr c_{2j} \right ) \; , \qquad
S_j^y = \frac{i}{2} \left ( c_{1j}^\dgr c_{2j} -
c_{2j}^\dgr c_{1j} \right ) \; ,
$$
\be
\label{16}
S_j^z = \frac{1}{2} \left ( c_{1j}^\dgr c_{2j} +
c_{2j}^\dgr c_{1j} \right ) \; ,
\ee
which gives
$$
c_{1j}^\dgr c_{1j} = \frac{1}{2} + S_j^x \; , \qquad
c_{2j}^\dgr c_{2j} = \frac{1}{2} - S_j^x \; ,
$$
\be
\label{17}
c_{1j}^\dgr c_{2j} = S_j^z - i S_j^y \; , \qquad
c_{2j}^\dgr c_{1j} = S_j^z + i S_j^y \; .
\ee

The physical meaning of the pseudospin operators (16) can be clarified by
introducing the left, $c_{jL}$, and the right, $c_{jR}$, location operators
\be
\label{18}
c_{jL} \equiv \frac{1}{\sqrt{2}} \; ( c_{1j} + c_{2j} ) \; , \qquad
c_{jR} \equiv \frac{1}{\sqrt{2}} \; ( c_{1j} - c_{2j} ) \; ,
\ee
characterizing the left or right location of an atom in the left or right 
well of a double-well potential. The pseudospin operators (16), expressed 
through the location operators (18), become
$$
S_j^x = \frac{1}{2} \left ( c_{jL}^\dgr c_{jR} +
c_{jR}^\dgr c_{jL} \right ) \; , \qquad
S_j^y = -\; \frac{i}{2} \left ( c_{jL}^\dgr c_{jR} -
c_{jR}^\dgr c_{jL} \right ) \; ,
$$
\be
\label{19}
S_j^z = \frac{1}{2} \left ( c_{jL}^\dgr c_{jL} -
c_{jR}^\dgr c_{jR} \right ) \; .
\ee
This representation demonstrates that $S_j^x$ characterizes the tunneling
intensity between the left and right wells of a double-well potential
centered at the $j$-site; the operator $S_j^y$ corresponds to the Josephson
current between the wells; while $S_j^z$ is the displacement operator
describing imbalance between the wells.

Finally, for Hamiltonian(10), we obtain the pseudospin form
\be
\label{20}
\hat H = E_0 N \; + \; \frac{1}{2}\;
\sum_{i\neq j} A_{ij} \; - \; \Om \sum_j S_j^x \; + \;
\sum_{i\neq j} B_{ij} S_i^x S_j^x \; - \;
\sum_{i\neq j} I_{ij} S_i^z S_j^z \; .
\ee
The first two terms here are not of the operator type, hence, can be
omitted. The third term describes the tunneling between the wells of a
double-well potential at the $j$-site, with the tunneling frequency $\Om$.
The fourth and fifth terms characterize particle interactions, with the
transverse strength $B_{ij}$ and longitudinal strength $I_{ij}$. The values
of the system parameters depend on the properties of particle interactions
and on the features of the lattice potential. Generally, these parameters
can be varied in rather wide ranges.

In order to illustrate how the tunneling frequency (15) can be varied,
we may take the double-well potential, in the vicinity of the lattice site
$\ba_j=0$, in the form
$$
V(\br) \simeq V_0 \left ( \frac{r_x}{r_0} \right )^2
\left [ \left ( \frac{r_x}{r_0} \right )^2 - 2 \right ] \; +
\; V_H(r_y,r_z) \; ,
$$
where $V_H(r_y,r_z)$ is a harmonic potential in the $y$- and $z$-directions.
The tunneling frequency $\Om$ essentially depends on the parameters of the
double-well potential $V(\br)$, its depth $V_0$ and the interwell distance
$r_0$. These parameters enter the variable
$$
\al \equiv \frac{1}{\sqrt{2m r_0^2 V_0} } \; ,
$$
which varies in the interval $0<\al<\infty$. For intermediate values of
$\al$, the tunneling frequency can be calculated numerically, while for
small and large $\al$, asymptotic expressions are available [36] yielding
$$
\Om \simeq 6 V_0 \exp\left ( -\; \frac{2}{\al} \right )
\qquad (\al \ll 1) \; ,
$$
$$
\Om \simeq V_0 \al^{2/3} \qquad (\al \gg 1) \; .
$$
These expressions show that the tunneling frequency $\Om$ can be regulated,
by changing $r_0$ and $V_0$, in a wide range $0<\Om<\infty$. This means
that varying the shape of the double-well potential, which is achievable
in experiments [16--21], one can regulate the tunneling properties of the
system.

\section{Equilibrium Phases}

Thermodynamics of the system with Hamiltonian (20), for sufficiently
long-range interactions, can be accurately described in the mean-field
approximation [37] corresponding to the equality
\be
\label{21}
S_i^\al S_j^\bt \; = \; < S_i^\al > S_j^\bt +
S_i^\al < S_j^\bt> - < S_i^\al><S_j^\bt> \qquad (i\neq j) \; ,
\ee
where the angle brackets imply statistical averaging. We introduce the
notation for the average interaction strengths
\be
\label{22}
A \equiv \frac{1}{N_L} \; \sum_{i\neq j} A_{ij} \; , \qquad
B \equiv \frac{1}{N_L} \; \sum_{i\neq j} B_{ij} \; , \qquad
I \equiv \frac{1}{N_L} \; \sum_{i\neq j} I_{ij} \; .
\ee
Also, we define the effective tunneling frequency
\be
\label{23}
\Om_{eff} \equiv \Om - 2B < S_j^x >
\ee
and the effective mean field
\be
\label{24}
h_{eff} \equiv \sqrt{ \Om_{eff}^2 + 4 I^2 < S_j^z>^2 } \; .
\ee

Using the standard methods of dealing with pseudospin Hamiltonians [38],
we find
$$
< S_j^x > \; = \; \frac{\Om_{eff}}{2h_{eff}}\;
{\rm tanh} \left ( \frac{h_{eff}}{2T} \right ) \; , \qquad
< S_j^y > \; = \; 0 \; ,
$$
\be
\label{25}
< S_j^z > \; = \; < S_j^z> \; \frac{I}{h_{eff}}\; {\rm tanh}
\left ( \frac{h_{eff}}{2T} \right ) \; ,
\ee
where $T$ is temperature. The system free energy reads as
\be
\label{26}
F = N E_0 + \frac{N}{2} \left ( A - 2B < S_j^x>^2 +
2I <S_j^z>^2 \right ) - NT \; \ln \left ( 2 {\rm cosh}\;
\frac{h_{eff}}{2T} \right ) \; ,
\ee
in which $N=N_L$ because of the unity filling factor. From the latter
equation, one can calculate all thermodynamics characteristics.

We shall be mainly interested in the properties of the average tunneling
intensity
\be
\label{27}
x \equiv \frac{2}{N_L} \; \sum_j < S_j^x > \; ,
\ee
average Josephson current
\be
\label{28}
y \equiv \frac{2}{N_L} \; \sum_j < S_j^y > \; ,
\ee
and the average well imbalance
\be
\label{29}
z \equiv \frac{2}{N_L} \; \sum_j < S_j^z > \; .
\ee

The properties of the system depend on the parameters defined in Eq. (22), 
which characterize the effective interaction strength. The parameter $A$ 
enters the free energy (26) additively, hence, it does not play here an 
important role. The parameter $B$ is composed of the effective interactions 
$B_{ij}$ given in Eq. (14). The symmetric terms $V_{ij}^{1111}$, 
$V_{ij}^{2222}$, and $V_{ij}^{1221}$ are close to each other, because 
of which the transverse interactions $B_{ij}$ are small, so that the 
parameter $B$ is substantially smaller than the parameter $I$. The main 
role among interactions is played by the longitudinal interaction strength 
$I$. Here we assume that $I$ is positive, which is the often situation for 
the exchange interactions [30,37,38]. Then, as follows from Hamiltonian 
(20), the nearest spins $S_i^z$ and $S_j^z$ tend to align parallel in 
order to reduce the system energy. Therefore only the ferromagnetic-type 
order is possible in the system [30,37,38].

Generally, the exchange interactions $I_{ij}$ could be negative. Then the 
longitudinal term in Hamiltonian (20) would enter with the sign plus, which 
would imply that the nearest spins prefer to align antiparallel to each 
other in order to lower the system energy. In that case, the 
antiferromagnetic-type order could be the sole possibility. In the present 
paper, we limit the consideration by a positive parameter $I>0$, which means 
that only a ferromagnetic-type order can arise.

To simplify the formulas, we use the dimensionless quantities, such as the
dimensionless tunneling frequency
\be
\label{30}
\om \equiv \frac{\Om}{I+B}
\ee
and the dimensionless transverse interaction
\be
\label{31}
b \equiv \frac{B}{I+B}  \; .
\ee
Also, we define the dimensionless field
\be
\label{32}
h \equiv \frac{h_{eff}}{I+B} \; ,
\ee
which, with the use of the above notations, becomes
\be
\label{33}
h =\sqrt{(\om-bx)^2 + (1-b)^2 z^2} \; .
\ee
>From definitions (27) to (29) and Eqs. (25), we have the tunneling intensity
\be
\label{34}
x =\frac{\om-bx}{h} \; {\rm tanh}
\left ( \frac{h}{2T} \right ) \; ,
\ee
Josephson current in equilibrium
\be
\label{35}
y = 0 \; ,
\ee
and the well imbalance
\be
\label{36}
z = z \; \frac{1-b}{h} \; {\rm tanh}
\left ( \frac{h}{2T} \right ) \; ,
\ee
where temperature $T$ is measured in units of $I+B$.

We may notice that Eqs. (34) to (36) are invariant under the replacement
\be
\label{37}
x \; \ra \; -x \; , \qquad \om \; \ra \; -\om \; , \qquad
z \; \ra \; -z \; .
\ee
Therefore, without the loss of generality, we can consider only the case
with $x\geq 0$, $\om\geq 0$, and $z\geq 0$.

Equation (36) shows that there can exist two thermodynamic phases, ordered
and disordered, when, respectively,
$$
z\neq 0 \qquad (ordered) \; ,
$$
$$
z = 0 \qquad (disordered) \; .
$$

In the ordered phase, the mean well imbalance is nonzero, and one has
\be
\label{38}
x = \om \qquad ( z\neq 0) \; ,
\ee
with $z$ given by the equation
\be
\label{39}
\frac{1-b}{h} \; {\rm tanh}
\left ( \frac{h}{2T} \right ) = 1 \; .
\ee
>From Eqs. (33) and (38), it follows that
$$
h = (1 - b) \sqrt{\om^2 + z^2 } \; .
$$
The ordered phase exists, when two conditions are valid, the tunneling
frequency is not too high,
\be
\label{40}
0 \leq \om < 1 \; ,
\ee
and the temperature is lower than the critical temperature
\be
\label{41}
T_c = \frac{(1-b)\om}{2\; {\rm artanh}\; \om} \; ,
\ee
so that
\be
\label{42}
0 \leq T < T_c \; .
\ee
When at least one of conditions (40) or (42) is not valid, that is, either
$\om\geq 1$ or $T\geq T_c$, the ordered phase cannot exist and transfers
to the disordered phase.

In the disordered phase, the mean well imbalance is zero, and one has
\be
\label{43}
x = {\rm tanh} \left ( \frac{\om-bx}{2T} \right ) \qquad
( z = 0 ) \; ,
\ee
which defines the tunneling intensity $x$.

At zero temperature $T=0$, the ordered phase is described by the equations
\be
\label{44}
x = \om \; , \qquad z = \sqrt{1-\om^2} \qquad (\om < 1 ) \; .
\ee
While the disordered phase is characterized by the expressions
\be
\label{45}
x = 1 \; , \qquad z = 0 \qquad (\om > 1) \; .
\ee
The quantum phase transition between the ordered and disordered phases
occurs at the tunneling frequency $\om=1$.

\section{Collective Excitations}

Collective excitations in the system represented by the pseudospin
Hamiltonian (20), correspond to pseudospin waves. The Heisenberg equations
of motion for the pseudospin operators yield
$$
\frac{dS_i^x}{dt} = 2 S_i^y \sum_{j(\neq i)} I_{ij} S_j^z \; ,
$$
$$
\frac{dS_i^y}{dt} = \Om S_i^z - 2 S_i^x
\sum_{j(\neq i)} I_{ij} S_j^z \; - \;
2 S_i^z \sum_{j(\neq i)} B_{ij} S_j^x \; ,
$$
\be
\label{46}
\frac{dS_i^z}{dt} = - \Om S_i^y + 2 S_i^y
\sum_{j(\neq i)} B_{ij} S_j^x \; .
\ee

For describing collective excitations, we employ the Fourier transformation
for the pseudospin operators
\be
\label{47}
S_j^\al = \frac{1}{N_L} \;
\sum_k \sgm_k^\al e^{i\bk\cdot\ba_j} \; , \qquad
\sgm_k^\al =  \sum_j S_j^\al e^{-i\bk\cdot\ba_j} \; ,
\ee
where $\bk$ pertains to the first Brillouin zone. Similarly, we expand the
interaction functions
\be
\label{48}
I_{ij} = \frac{1}{N_L} \;
\sum_k I_k e^{i\bk\cdot\ba_{ij}} \; , \qquad
B_{ij} = \frac{1}{N_L} \;
\sum_k B_k e^{i\bk\cdot\ba_{ij}} \; ,
\ee
in which $\ba_{ij}=\ba_i-\ba_j$. Then, Eqs. (46) acquire the form
$$
\frac{d\sgm_k^x}{dt} = \frac{2}{N_L} \;
\sum_p I_{k-p} \sgm_p^y \sgm_{k-p}^z \; ,
$$
$$
\frac{d\sgm_k^y}{dt} = \Om \sgm_k^z \; - \; \frac{2}{N_L} \;
\sum_p I_{k-p} \sgm_p^x \sgm_{k-p}^z \; - \; \frac{2}{N_L} \;
\sum_p B_{k-p} \sgm_p^z \sgm_{k-p}^x \; ,
$$
\be
\label{49}
\frac{d\sgm_k^z}{dt} = - \Om \sgm_k^y \; + \; \frac{2}{N_L} \;
\sum_p B_{k-p} \sgm_p^y \sgm_{k-p}^x \; ,
\ee
where $\sgm_k^\al=\sgm_k^\al(t)$.

To find the spectrum of collective excitations, we resort to the
random-phase approximation. For this purpose, we look for the solutions
to Eqs. (49) in the form
\be
\label{50}
\sgm_k^\al(t) \; = \; < \sgm_k^\al> + \;
\dlt \sgm_k^\al e^{-i\ep t} \; ,
\ee
describing the deviations from the equilibrium values
$$
< \sgm_k^\al > \; = \; \dlt_{k0} N_L < S_j^\al > \; .
$$
Linearizing Eqs. (49) with respect to $\dlt\sgm_k^\al$ yields for the
spectrum of pseudospin waves
\be
\label{51}
\ep_k^2 =  ( \Om - Bx) [ \Om - (B + I_k ) x ] +
I ( I + B_k ) z^2 \; .
\ee
In deriving Eq. (51), we have taken into account that
$$
\lim_{k\ra 0} I_k =  I \; , \qquad
\lim_{k\ra 0} B_k = B \; ,
$$
with $I$ and $B$ defined in Eq. (22).

To consider the spectrum property in the long-wave limit, we use the
asymptotic equalities
\be
\label{52}
I_k \simeq I \; - \; \frac{1}{2} \;
\sum_i I_{ij} (\bk\cdot \ba_{ij} )^2 \; , \qquad
B_k \simeq B \; - \; \frac{1}{2} \;
\sum_i B_{ij} (\bk\cdot \ba_{ij} )^2 \; ,
\ee
where $k\ra 0$. We define the gap $\Dlt$ by the equation
\be
\label{53}
\left ( \frac{\Dlt}{I+B} \right )^2 =
( \om - bx) ( \om - x) + ( 1 - b) z^2 \; .
\ee
And let us introduce a matrix $G_{ij}$, given by the equality
\be
\label{54}
\frac{G_{ij}}{I+B} =  ( \om - bx ) x I_{ij} -
( 1 - b) z^2 B_{ij} \; .
\ee
Then the long-wave limit of spectrum in Eq. (51) gives
\be
\label{55}
\ep_k^2 \simeq \Dlt^2 \; + \; \frac{1}{2} \;
\sum_i G_{ij} (\bk\cdot \ba_{ij} )^2 \; .
\ee
The spectrum is quadratic, which is typical of spin waves.

Specifying the gap $\Dlt$ and the matrix $G_{ij}$ for the ordered and
disordered phases, we have for the ordered phase, when $z\neq 0$,
$$
\frac{\Dlt}{I+B} = \sqrt{1-b}\; z \; , \qquad
\frac{G_{ij}}{I+B} = ( 1 - b)
\left ( \om^2 I_{ij} - z^2 B_{ij} \right ) \; ,
$$
and for the disordered phase, with $z=0$,
$$
\frac{\Dlt}{I+B} = \sqrt{ ( \om - bx) ( \om -x) } \; , \qquad
\frac{G_{ij}}{I+B} = ( \om - bx) x I_{ij} \; .
$$
The gap disappears at the critical point of the order-disorder phase
transition. Approaching this point from the side of the ordered phase, one
has $z\ra 0$, while approaching the point from the side of the disordered
phase, one has $x\ra\om$. In both these cases, $\Dlt\ra 0$ at the critical
point.

\section{Nonequilibrium Loading}

Let us suppose that atoms are loaded into a double-well lattice in an
initially nonequlibrium state. For describing their equilibration process,
one needs to study the temporal behavior of the variables defined in
Eqs. (27) to (29) and characterizing the tunneling intensity $x=x(t)$,
Josephson current $y=y(t)$, and the well imbalance $z=z(t)$. The evolution
equations for these quantities can be obtained by averaging the operator
equations (46). For this purpose, one often employs the mean-field
approximation. This approximation does not take into account the relaxation
processes due to atomic interactions. However, such processes could be
important in order that equilibrium would be possible. Therefore, we shall
resort to a more accurate approximation, called the local-field approximation.
This approximation, suggested by Wangness [39], treats atomic interactions
as occurring in a local field formed by other particles, which results in
the appearance of relaxation characterized by the attenuation parameters,
$\gm_1$ and $\gm_2$ whose values can be calculated through the given atomic
interactions. The local-field approximation assumes the existence of local
equilibrium [40], so that at each moment of time the variables tend to relax
to the locally-equilibrium state. Averaging Eqs. (46) in the frame of the
local-field approximation [39] yields the system of equations
$$
\frac{dx}{dt} = ( 1 - b) yz - \gm_2 ( x - x_t) \; , \qquad
\frac{dy}{dt} = (\om - x ) z - \gm_2 ( y - y_t) \; ,
$$
\be
\label{56}
\frac{dz}{dt} = ( bx - \om) y - \gm_1 ( z - z_t ) \; ,
\ee
where the time variable is measured in units of $1/(I+B)$, while the 
attenuation parameters $\gm_1$ and $\gm_2$, in units of $I+B$, and the 
local fields are
$$
x_t = \frac{\om-bx}{h} \;
{\rm tanh} \left ( \frac{h}{2T} \right ) \; ,
\qquad y_t = 0 \; ,
$$
\be
\label{57}
z_t = \frac{1-b}{h} \; z\;
{\rm tanh} \left ( \frac{h}{2T} \right ) \; .
\ee
Here $T$ is the temperature, in units of $I+B$, corresponding to a would
be equilibrium state, the type of the latter being defined by the system
parameters. The explicit expression for $h$ is given in Eq. (33). The
evolution equations (56) are supplemented with the initial conditions
\be
\label{58}
x(0) = x_0 \; , \qquad y(0)=y_0 \; , \qquad z(0)=z_0 \; ,
\ee
which, generally, represent a nonequilibrium state.

Keeping in mind applications to ultracold atoms, we may set temperature
to zero. Then, Eqs. (56) possess two types of stationary solutions. One
type is defined by the fixed point
\be
\label{59}
x_1^* = \om \; , \qquad y_1^* = 0 \; , \qquad
z_1^* =\sqrt{1-\om^2} \; ,
\ee
which corresponds to the ordered phase. While another type is given by the
fixed point
\be
\label{60}
x_2^* = 1 \; , \qquad y_2^* = 0 \; , \qquad
z_2^* = 0 \; ,
\ee
corresponding to the disordered phase. The question remains whether
an atomic system loaded into the double-well lattice, in an initially
nonequilibrium state, would relax to one of the stationary solutions.

\section{Stability Analysis}

The stability of stationary solutions can be studied by the Lyapunov
stability analysis. To this end, we calculate, in the standard way, the
Jacobian matrix associated with the dynamical system (56). Then  we find
the eigenvalues of the Jacobian matrix, denoted by $\lbd$, which give the
characteristic exponents. The real parts ${\rm Re}\; \lbd$ define the
Lyapunov exponents describing the stability properties. For simplicity,
we set $\gm_1=\gm_2=\gm$.

Accomplishing this procedure for the fixed point (59), we find that it
is stable when $\om<1$. One of the characteristic exponents is exactly
\be
\label{61}
\lbd_1 = -\gm \; .
\ee
The expressions for two other characteristic exponents are rather cumbersome,
because of which we write them down to the first order in $\gm$, resulting in
\be
\label{62}
\lbd_{2,3} \simeq -\; \frac{2-b-\om^2}{2(1-b)} \;
\gm \pm i\om_{eff} \; ,
\ee
with the effective frequency
\be
\label{63}
\om_{eff} =\sqrt{(1-b)\left ( 1-\om^2\right ) }
\qquad (\om < 1 ) \; .
\ee
This shows that the fixed point (59) is a stable focus, when $\om<1$ and
$\gm$ is small. Recall that by its definition (31), one always has $b<1$.

The fixed point (60) is stable for $\om>1$. Then, one of the characteristic
exponents is exactly the same as Eq. (61). Two other characteristic exponents,
to the first order in $\gm$, are
\be
\label{64}
\lbd_{2,3} \simeq -\; \frac{2\om-1-b}{2(\om-b)} \;
\gm \pm i\om'_{eff} \; ,
\ee
with the effective frequency
\be
\label{65}
\om'_{eff} =\sqrt{(\om -1)(\om - b ) }
\qquad (\om > 1 ) \; .
\ee
Thus, the fixed point (60) is also a stable focus for $\om>1$ and small
$\gm$.

The value of the tunneling frequency $\om=1$ is a bifurcation point
related to the dynamical phase transition. At this point, the dynamical
system is neutral, with the characteristic exponents
$$
\lbd_1= -\gm \; , \qquad \lbd_2 = 0 \; , \qquad
\lbd_3= -\gm \qquad (\om=1)
$$
for any $\gm$.

An important observation, following from the above analysis, is that the
dynamical system is structurally stable only in the presence of the damping
parameter $\gm>0$. Setting the latter to zero leads to the characteristic
exponents for the ordered fixed point (59)
$$
\lbd_1 = 0 \; , \qquad \lbd_{2,3} =\pm i\om_{eff} \qquad
(\gm=0,\; \om < 1)
$$
and for the disordered fixed point (60)
$$
\lbd_1 = 0 \; , \qquad \lbd_{2,3} =\pm i\om'_{eff} \qquad
(\gm=0,\; \om > 1)
$$
where $\om_{eff}$ and $\om'_{eff}$ are given by Eqs. (63) and (65).
These characteristic exponents demonstrate that the dynamical system is
structurally unstable.

But the case of no attenuation corresponds to the usual mean-field
approximation. Hence the above consideration shows that the mean-field
approximation results in a structurally unstable dynamical system, thus,
this approximation cannot correctly describe the evolutional processes
of atoms in a double-well lattice.

In order to illustrate in an explicit way the difference between the
temporal behavior of solutions to the evolution equations (56) and the
related mean-field approximations, with no attenuation, we solve Eqs. (56)
numerically. The initial conditions (58) are chosen so that to correspond
to an initially nonequilibrium state.

In Fig. 1, the relaxation to the ordered stationary solution (59) is shown
for $\gm=1$ and $\om<1$. Taking the attenuation parameter $\gm$ equal to 
one in dimensionless units implies that it is of order $I+B$ in dimensional 
units. For these parameters, the fixed point (59) becomes a stable node. 
Contrary to this, the mean-field approximation, with $\gm=0$, exhibits 
permanently oscillating solutions. The curves for $x(t)$ and $z(t)$ always 
oscillate around the given initial conditions, which is not a correct
behavior. In turn, the latter is caused by the structural instability of 
the dynamical system in the mean-field approximation. The relaxation to the
disordered stationary solution (60) is demonstrated in Fig. 2 for $\om>1$
and $\gm=1$, when Eq. (60) corresponds to a stable node. But the mean-field
approximation, with $\gm=0$, again exhibits an incorrect oscillatory
behavior depending on the choice of the initial conditions (58). Again, 
it is the structural instability that is responsible for the incorrect 
dynamics in the mean-field approximation.

Full equations (56) can also exhibit the oscillatory behavior of their
solutions. This happens when the fixed points are the stable focuses, which
occurs for $\gm\ll\om$, as follows from Eqs. (62) to (65). In particular, if
$\om\gg 1$, than, according to Eq. (65), the oscillation period approximately
is $2\pi/\om$. This situation is illustrated in Fig. 3.

\section{Discussion}

We have considered a system of atoms in a double-well optical lattice.
The case of an insulating lattice with the unity filling factor is studied.
This type of double-well optical lattices is of special interest, being a
convenient setup for realizing the ordered and disordered states of atoms 
in a double-well lattice.

The principal difference of the present paper, as compared to the earlier 
theoretical works on the double-well lattices [22--24,41], where the 
superfluidity-insulator phase transition is studied for atoms with local 
interactions, is that we consider an insulating double-well lattice, with 
atoms possessing long-range interactions. An effective Hamiltonian in the 
pseudospin representation is derived. The system exhibits two thermodynamic 
phases, ordered and disordered. The phase transition between the phases can 
be driven either by temperature or by varying the system parameters. For 
instance, one can vary the strength of atomic interactions or the shape 
of the double-well potential.

The spectrum of collective excitations has the form typical of that for
pseudospin waves. In the long-wave limit, the spectrum is quadratic, with
a gap. The latter disappears at the phase transition point.

Nonequilibrium properties of the system are investigated. Physically, 
the situation can correspond to atoms loaded into a lattice in an 
initially nonequilibrium state. Such a nonequilibrium state can also 
be prepared by disturbing the lattice by external fields.

The relaxation of the system, from an initially nonequilibrium state 
to equilibrium, cannot be described by the mean-field approximation. 
This is rather clear from the physical point of view, since the 
mean-field approximation does not take into account attenuation effects, 
hence, is not able to describe the equilibration process in principle. 
>From the mathematical point of view, as we show, the mean-field 
approximation results in a structurally unstable dynamical system. 
To correctly describe the process of relaxation, the damping effects, 
caused by atomic interactions, must be included. This can be done, e.g., 
by employing the local-field approximation.

The long-range interactions that would be sufficient for realizing the 
effects described above could be of dipolar type, such that occur for 
polar molecules [31,42], Rydberg atoms [32], and quantum gases with dipolar 
atomic interactions (see review articles [33,43]). A good candidate would 
be the gas of cold atoms $^{52}$Cr, possessing large magnetic moments of 
6$\mu_B$ and, as a result, sufficiently strong dipolar interactions [44,45]. 
Because of the existence of several species exhibiting long-range dipolar 
intreations, the latter can be of different strength. In addition, these 
interactions can be effectively modulated by external magnetic or electric 
fields, thus, tuning the interaction strength in a wide range [46--49].

In the dynamics of atoms, relaxing from a nonequilibrium state to their 
equilibrium, an important role is played by the attenuation parameter 
$\gm$. The calculation of the latter, for systems with dipolar interactions, 
is a well known procedure, described in detail in literature on magnetic 
resonance [50--55]. In the case of particles with magnetic moment $\mu$, 
interacting through magnetic dipolar forces, an accurate value of the 
attenuation parameter is given by the expression $\gm=\rho z_0\mu^2/\hbar$, 
where $\rho$ is the mean particle density and $z_0$ is the number of nearest 
neighbors. This value of $\gm$ is of the order of the interaction strength. 
That is why setting, in dimensionless units, $\gm$ to one, as we have done 
in numerical calculations, is absolutely natural.

Taking, for illustration, the magnetic moment of $^{52}$Cr, equal to $\mu=
6\mu_B$, with $\mu_B=0.927\times 10^{-20}$ erg/G being the Bohr magneton, 
the typical atomic densities in a trap $\rho\sim(10^{12}-10^{15})$ 
cm$^{-3}$, and accepting for the number of nearest neighbors $z_0\sim 10$, 
we have $\gm\sim(1-10^3)$ s$^{-1}$. This means that the relaxation time 
$T_{rel}\equiv 1/\gm$ is of order $T_{rel}\sim(10^{-3}-1)$ s. The lifetime 
of trapped atomic systems can vary in the range of $t_{exp}\sim(1-100)$ s. 
Therefore, the relaxation time $T_{rel}$ in many cases is much smaller than 
$t_{exp}$, which implies that taking into account the attenuation is of 
critical importance. The simple mean-filed approximation without taking 
account of the relaxation effects would be qualitatively incorrect.

In conclusion, it is worth noting that the shape of the double-well 
potential can be easily regulated. Hence, it is possible to organize any 
required sequence of transitions between the ordered and disordered states 
in the double-well lattice during the system lifetime.

\vskip 5mm

{\bf Acknowledgement}

\vskip 2mm

We are grateful for financial support to the Russian Foundation for Basic
Research (Grant 08-02-00118).

\newpage

{\Large {\bf Appendix A}}

\vskip 5mm

The eigenproblem for the lattice Hamiltonian (2),
$$
H_L(\br) \vp_{nk}(\br) = E_{nk} \; \vp_{nk}(\br) \; ,
$$
defines the Bloch functions $\vp_{nk}(\br)$ and the Bloch spectrum
$E_{nk}$. This eigenproblem, invoking the relation between the Bloch and
Wannier functions
$$
\vp_{nk}(\br) = \frac{1}{\sqrt{N_L}} \;
\sum_j w_n(\br-\ba_j) \; e^{i\bk\cdot\ba_j} \; ,
$$
can be rewritten as
$$
H_L(\br) \; w_n(\br-\ba_j) = \frac{1}{N_L} \;
\sum_{ik} E_{nk} \; e^{i\bk\cdot\ba_{ij}} \;
w_n(\br-\ba_i) \; ,
$$
where $\ba_{ij}\equiv\ba_i-\ba_j$. Using this in Eq. (5) gives Eq. (6),
with
$$
E_{ij}^n = \frac{1}{N_L} \;
\sum_k E_{nk} \; e^{i\bk\cdot\ba_{ij}} \; .
$$
Representing the latter in form (7) yields
$$
E_n = \frac{1}{N_L} \; \sum_k E_{nk}
$$
and
$$
J_{ij}^n = \frac{1}{N_L} \; \sum_k E_{nk} \;
e^{i\bk\cdot\ba_{ij}}  \qquad (i\neq j) \; .
$$
These properties are used in deriving Hamiltonian (10).

\newpage

\begin{center}
{\large{\bf Figure Captions}}

\end{center}

\vskip 5mm

{\bf Fig. 1}. Dimensionless variables describing the tunneling
intensity $x(t)$, Josephson current $y(t)$, and well population
imbalance $z(t)$ as functions of dimensionless time for $\om=0.1$
and $b=0.5$. Initial conditions are $x_0=0.66$, $y_0=0.75$, and
$z_0=0$. The case of no attenuation $(\gm=0)$ is shown by the dashed
curve. The case with attenuation $(\gm=1)$ is represented by the
solid line.

\vskip 5mm

{\bf Fig. 2}. Dimensionless variables $x(t)$, $y(t)$, and $z(t)$ as
functions of dimensionless time for $\om=1.5$ and $b=0.5$. Initial
conditions are $x_0=0.33$, $y_0=0.5$, and $z_0=0.8$. The attenuation
parameters are: $\gm=0$ (dashed line) and $\gm=1$ (solid line).

\vskip 5mm

{\bf Fig. 3}. Population imbalance for large tunneling $\om=100$,
with $\gm=1$ and $b=0.5$ as a function of dimensionless time. Initial
conditions are $x_0=0$, $y_0=0$, and $z_0=1$.

\newpage

\begin{figure}[hbtp]
\vspace{9pt}
\centerline{
\hbox{ \includegraphics[width=8cm]{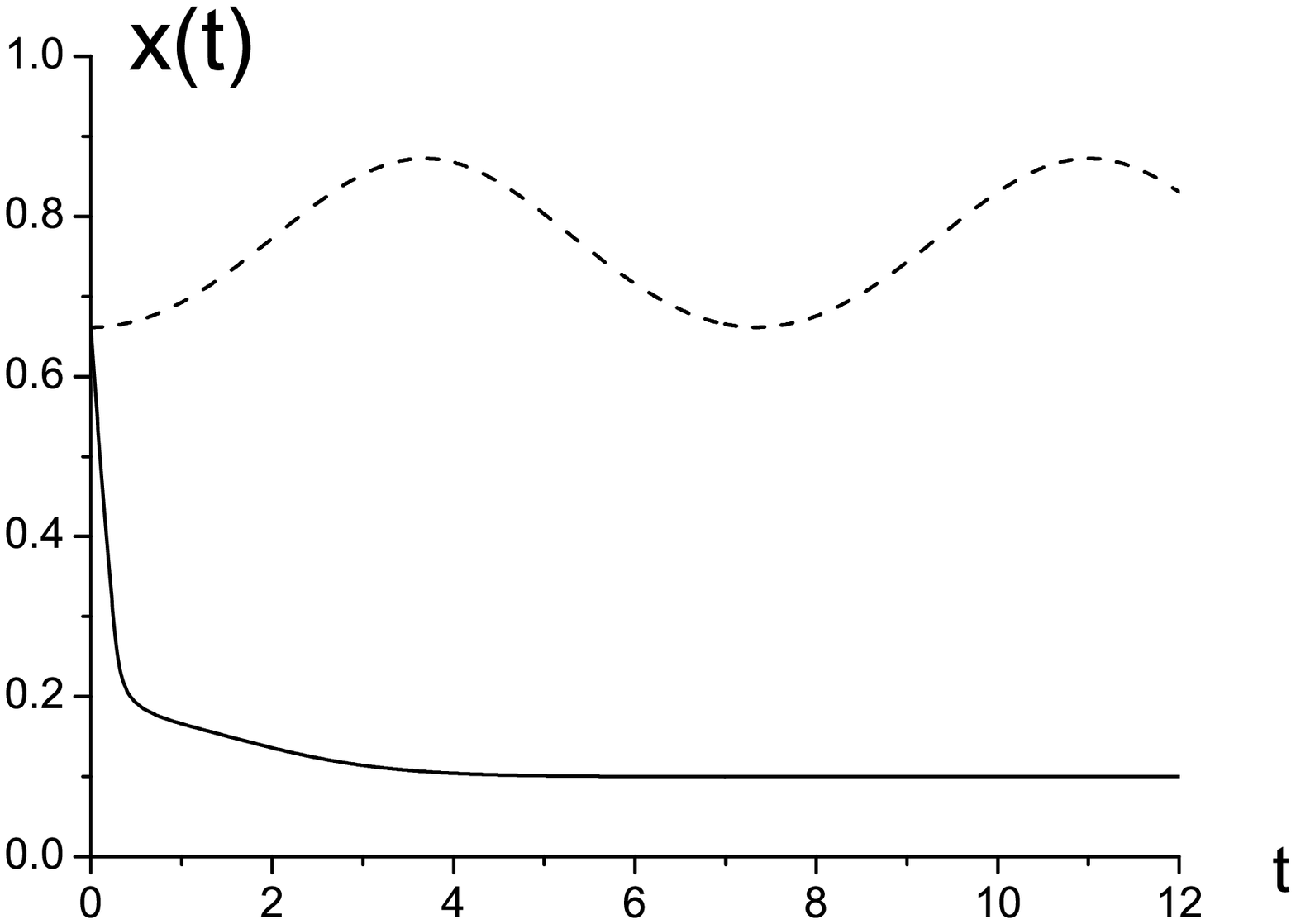} } }
\vspace{9pt}
\centerline{
\hbox{ \includegraphics[width=8cm]{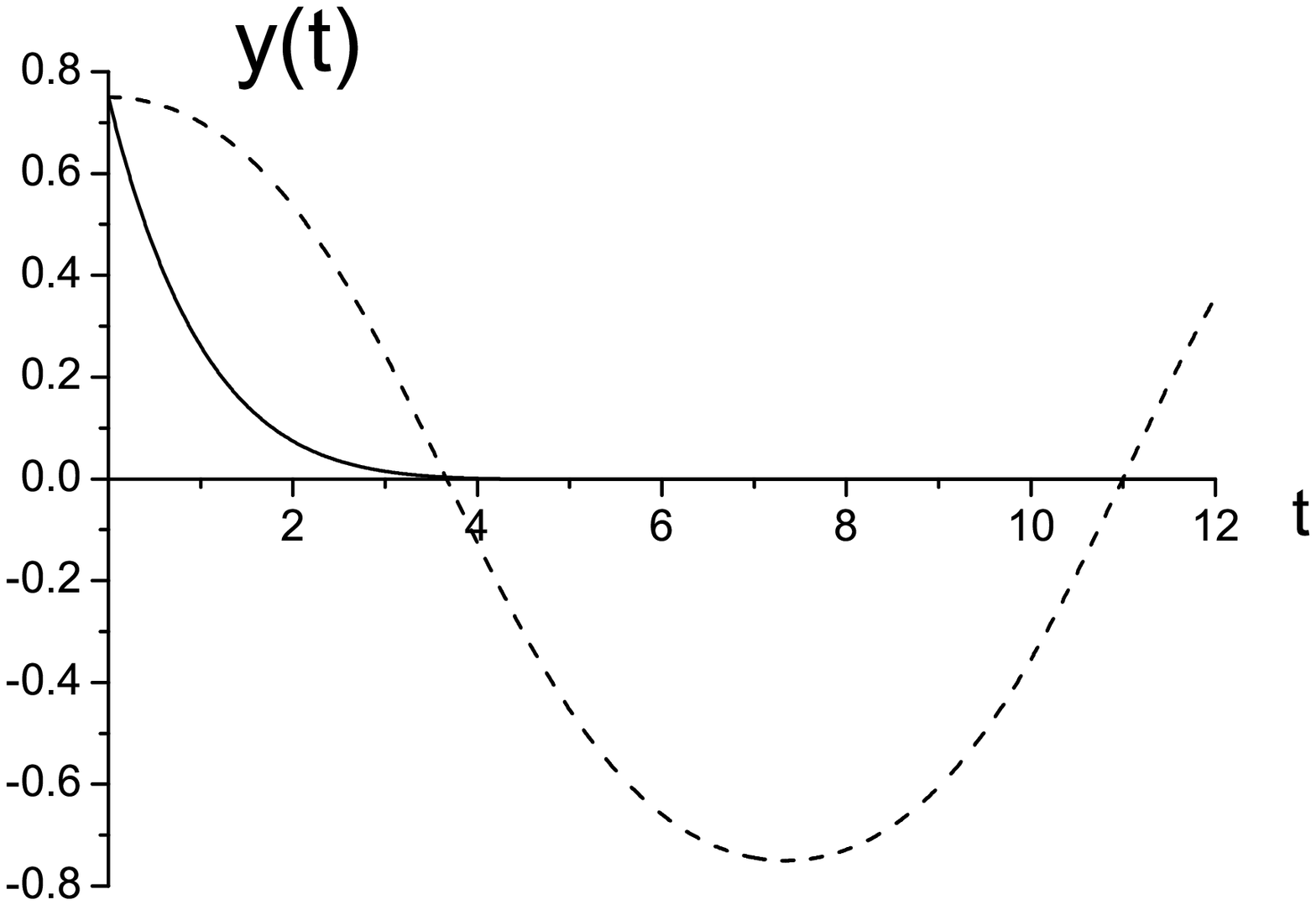} \hspace{1cm}
\includegraphics[width=8cm]{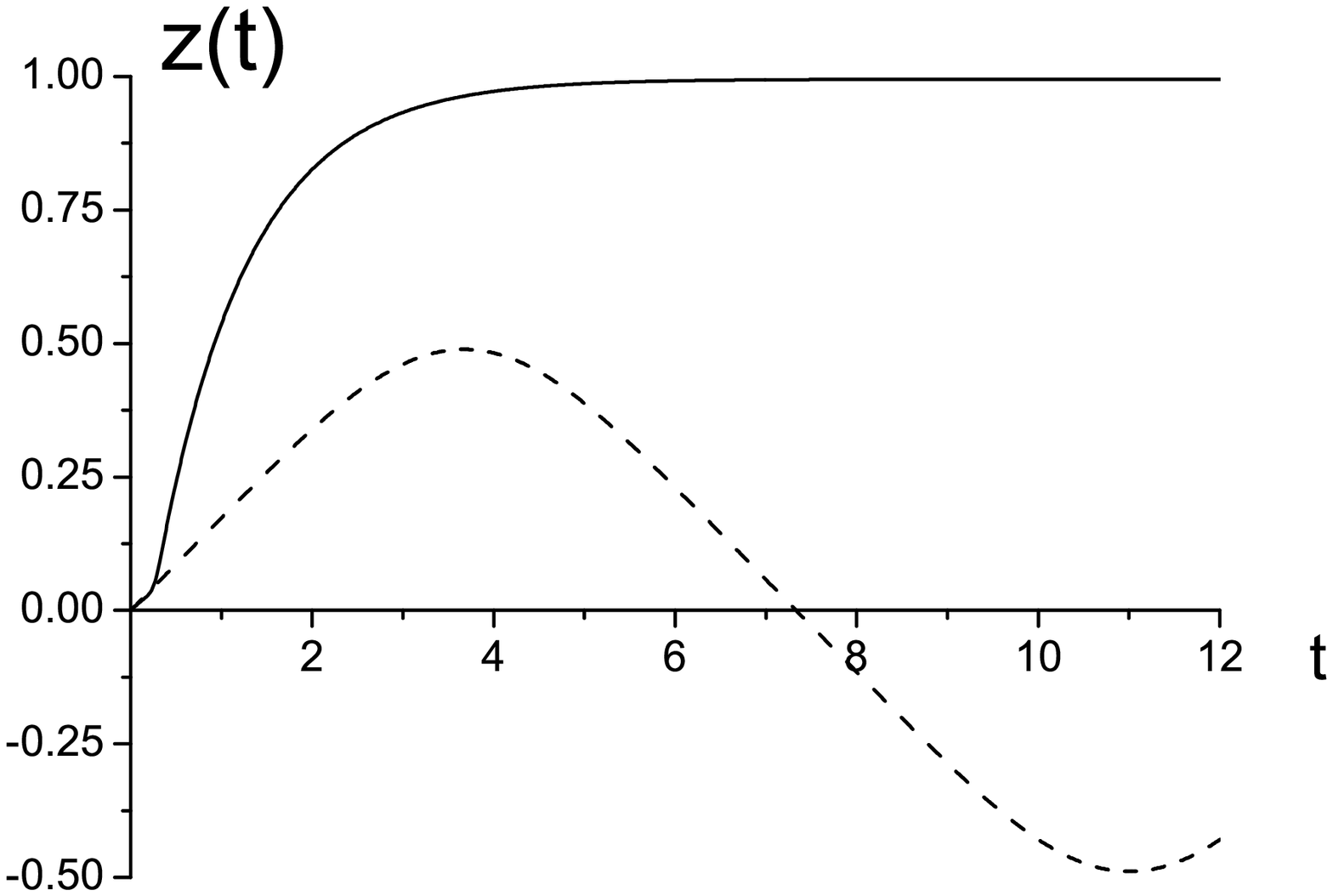} } }
\caption{Dimensionless variables describing the tunneling
intensity $x(t)$, Josephson current $y(t)$, and well population
imbalance $z(t)$ as functions of dimensionless time for $\om=0.1$
and $b=0.5$. Initial conditions are $x_0=0.66$, $y_0=0.75$, and
$z_0=0$. The case of no attenuation $(\gm=0)$ is shown by the dashed
curve. The case with attenuation $(\gm=1)$ is represented by the solid
line.}
\label{fig:Fig.1}
\end{figure}

\newpage

\begin{figure}[hbtp]
\vspace{9pt}
\centerline{
\hbox{ \includegraphics[width=8cm]{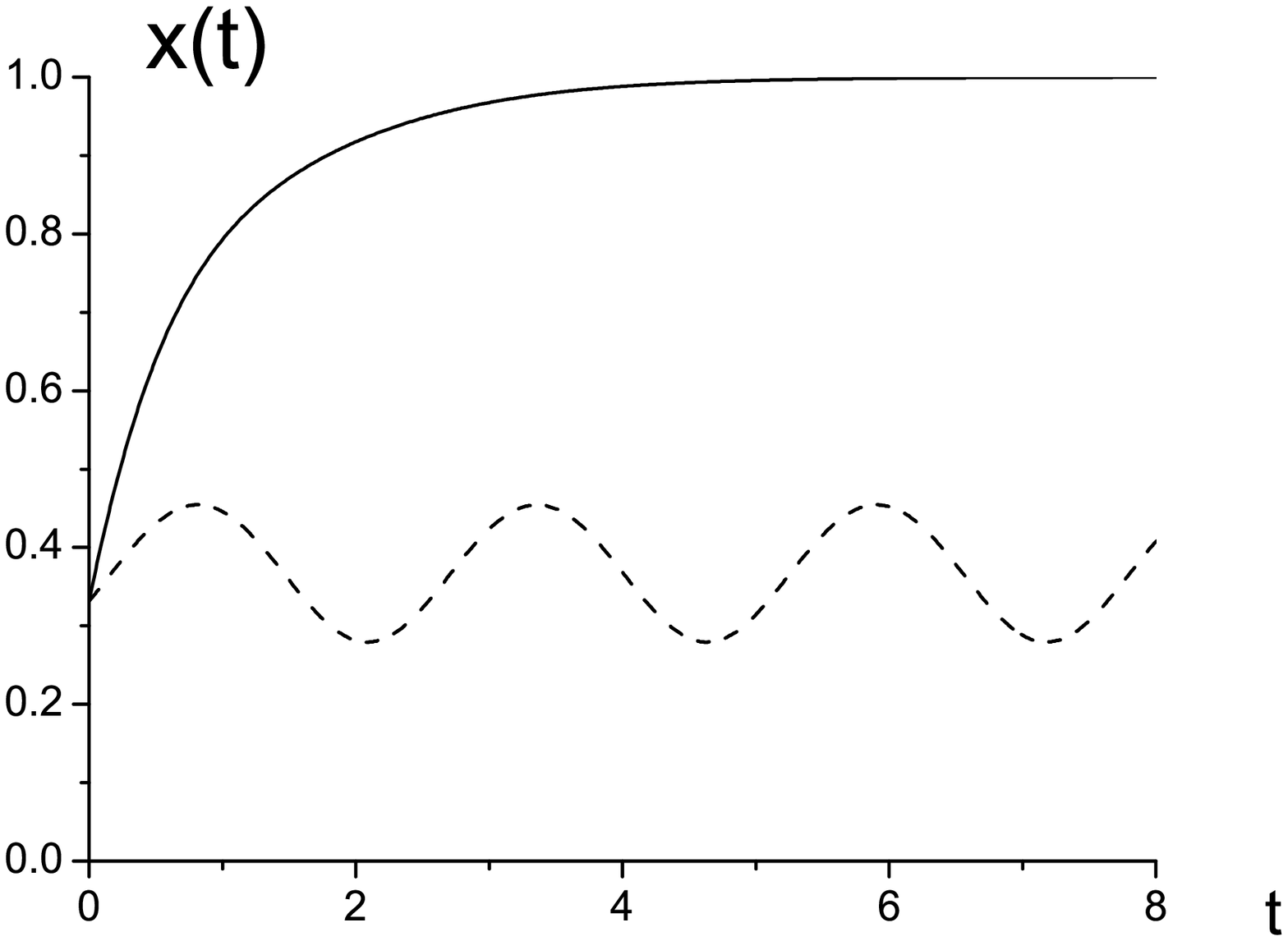} } }
\vspace{9pt}
\centerline{
\hbox{ \includegraphics[width=8cm]{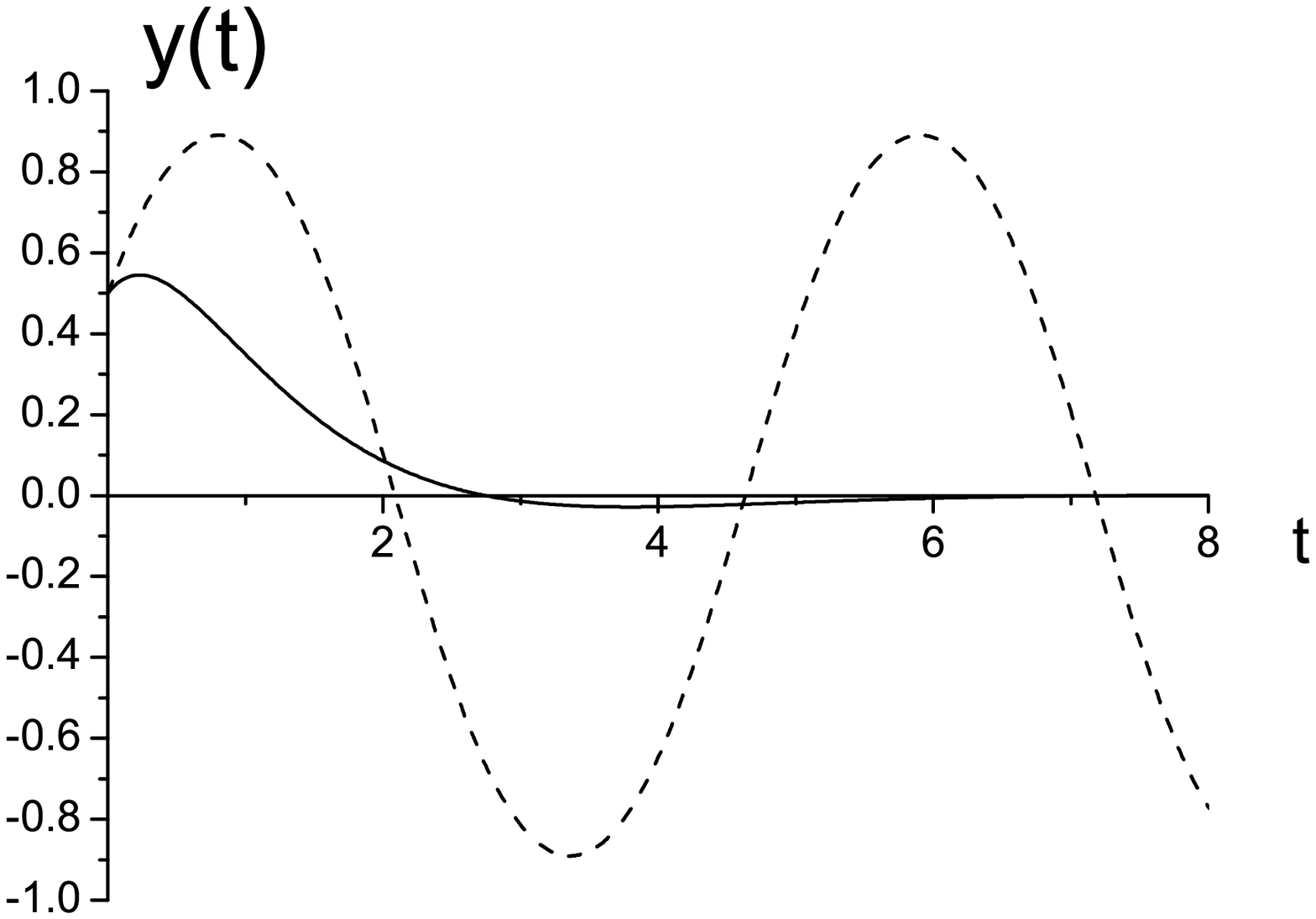} \hspace{1cm}
\includegraphics[width=8cm]{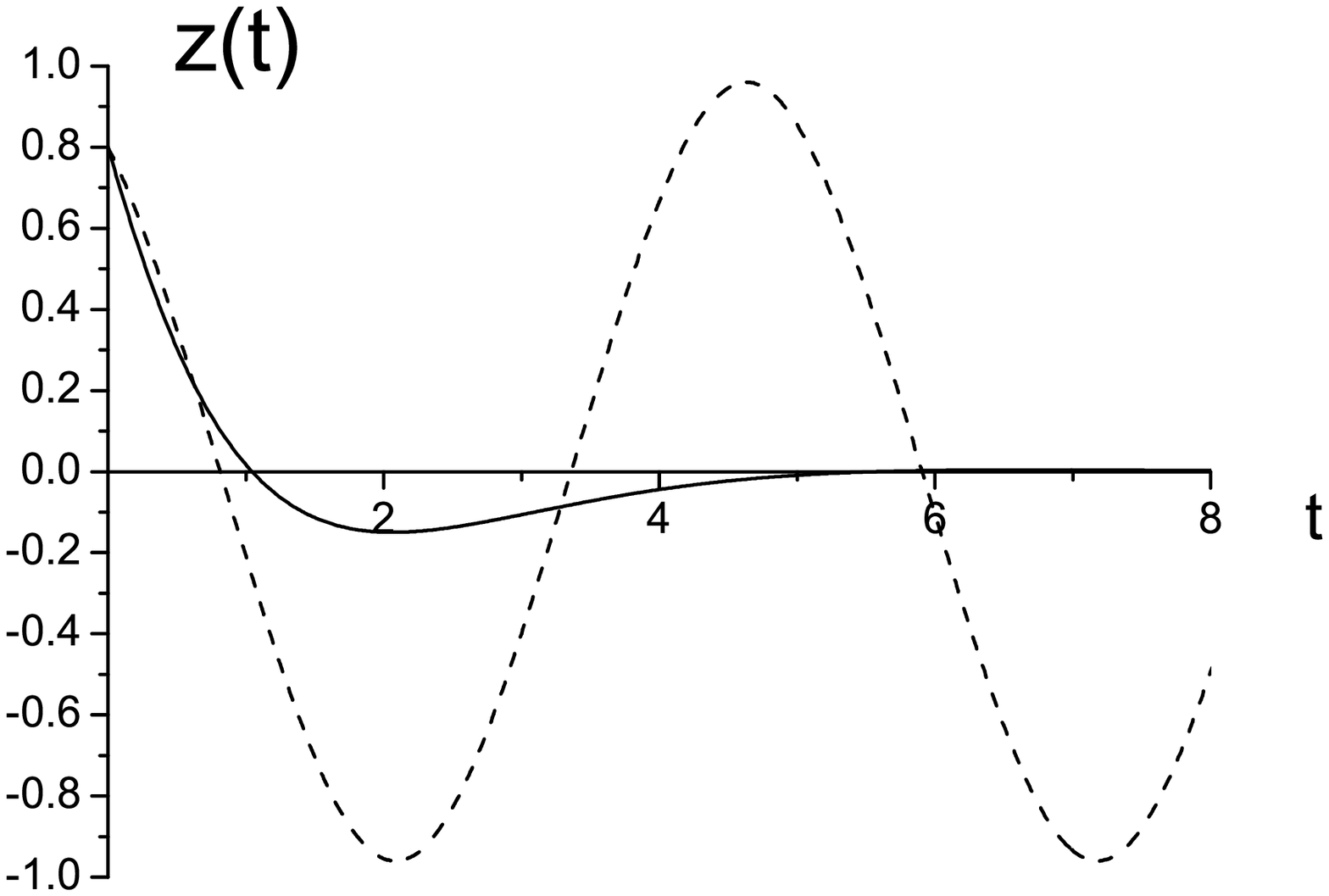} } }
\caption{Dimensionless variables $x(t)$, $y(t)$, and $z(t)$ as
functions of dimensionless time for $\om=1.5$ and $b=0.5$. Initial
conditions are $x_0=0.33$, $y_0=0.5$, and $z_0=0.8$. The attenuation
parameters are: $\gm=0$ (dashed line) and $\gm=1$ (solid line).}
\label{fig:Fig.2}
\end{figure}

\newpage

\begin{figure}[h]
\epsfig{figure=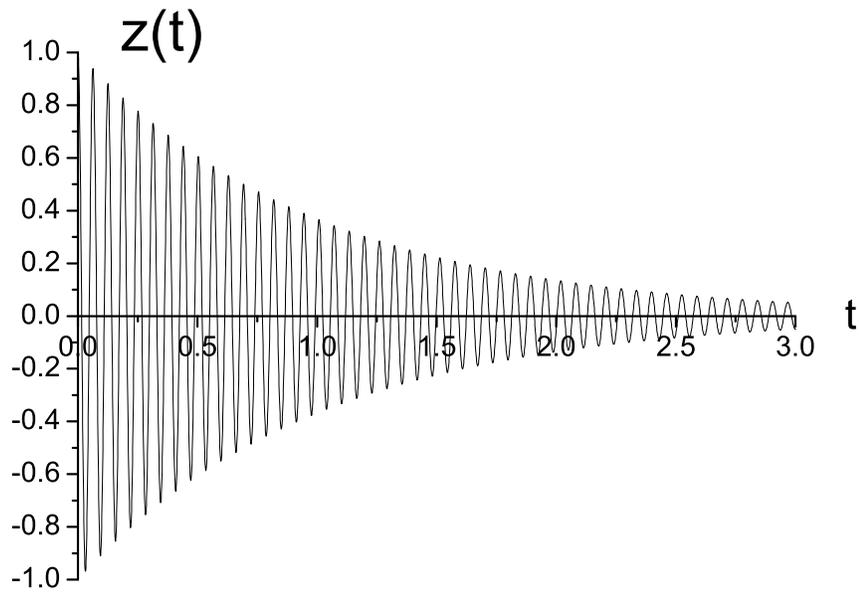,width=14cm}
\caption{Population imbalance for large tunneling $\om=100$,
with $\gm=1$ and $b=0.5$ as a function of dimensionless time. Initial
conditions are $x_0=0$, $y_0=0$, and $z_0=1$.
}
\label{Fig.3}
\end{figure}

\end{document}